\documentstyle[preprint,prl,aps]{revtex}

\begin{document}

%\draft

\title{Generalized quantum measurements and local realism}

\author{ Marek \.Zukowski \cite{poczta2},
Ryszard Horodecki \cite{poczta1}}

\address{Institute of Theoretical Physics and Astrophysics\\
University of Gda\'nsk, 80--952 Gda\'nsk, Poland}

\author{Micha\l{} Horodecki}

\address{Department of Mathematics and Physics\\
 University of Gda\'nsk, 80--952 Gda\'nsk, Poland}

\author{Pawe\l{} Horodecki}

\address{Department of Technical Physics and Applied Mathematic\\
Technical University of Gda\'nsk, 80--952 Gda\'nsk, Poland}

\maketitle

 \begin{abstract}
 The structure of a local hidden variable model for experiments involving
sequences of measurements  is analyzed. Constraints imposed by local
realism on the conditional probabilities of the outcomes of such
 measurement schemes are explicitly derived. The violation of
local realism in the case of ``hidden nonlocality''
is illustrated by an operational example.
 \end{abstract}
\pacs{}
\newpage
\section{Introduction}
The question of the possibility of the existence of a local realistic model
for quantum predictions for pure entangled
states has has found a negative answer.
Namely,
Gisin \cite{Gis1} proved
that the only pure states of two-component system which do not violate the
Bell-CHSH inequality are the product states (states of such a
property are often,
slightly misleadingly, called the ``local" ones).
These results, improved by Gisin and Peres \cite{Gis2}, have been
generalised by Rohrlich and Popescu \cite{Rohr} to N-component quantum
systems.

However in the case of mixed states the problem becomes much more complicated.
One might naively think that analogously to the case
of pure states, the only
mixed states which do not violate Bell's inequalities are the mixtures of
product states (i.e. separable states). In his pioneering paper
Werner \cite{Werner} showed that this
conjecture is false. He studied  the
possibility of a direct construction of a
local hidden variable (LHV) model for some families of	mixed states, and
showed that there is a class of nonseparable mixtures for which  the results
of performed measurements can be simulated
by such a model. However, in a recent development, Popescu \cite{pop1}
noticed that Werner
had considered only a restricted class of measurement procedures.
Namely, Werner
constructed a LHV model for single (i.e., nonsequencial)
von Neumann measurements. Later, again,  Popescu \cite{pop2} was the first to
show that most of the Werner mixtures exhibit violations
of local realism
if sequences of measurements are taken into account.
Such an exposure of the so-called hidden nonlocality \cite{pop2} involves
sequences of two measurements.
The  initial ensemble of two particle systems is subjected to the first
measurement. Afterwards, a
subensemble of the pairs which produced some required
outcome is selected and tested by measuring the Bell observable. If the
subensemble does not satisfy  Bell inequalities, then one concludes that the
original ensemble violates local realism.

Recently, Gisin \cite{Gis3} has shown that for two-level
systems the nonlocality can be revealed by using filters at the first stage of
the process
(a procedure of this kind
can be treated as generalized
measurement).
Quite recently,
Peres \cite{Peres} considered collective tests of  particles
in the Werner state and used consecutive measurements to show the
impossibility of constructing a local realistic description for some
processes of this kind \footnote{In fact, it has been proved, that
all the inseparable Werner states are nonlocal, by using a recursive
protocol in the process of the so-called distillability \protect \cite{dist}.
More generally, quite recently it is been shown \cite{distn} that any inseparable $2\times2$
system can be distilled.}.

In this paper we aim at a description of ``hidden nonlocality''
which is parallel to the original Bell approach. Namely, Bell simply showed
that some functions cannot be simulated by suitable averages of product ones.
It is rather obvious that for better understanding of a ``hidden nonlocality''
the latter should be presented in complete analogy to Bell's standard
nonlocality.

Rigorous proof that the local hidden variable
model
for {\it joint} probabilities of outcomes of sequences of measurements
cannot exist in the case of ``hidden nonlocality'' will be given
in section 2.
More precisely,  we will explicitly show that, if the
conditional probabilities cannot be simulated by suitable averages of
product of functions depending upon hidden variables
then also the joint probabilities of sequences of
outcomes cannot be represented in this way.
In section 3 we also provide a proposal of a
feasible experiment for which there is no
local-realistic description
despite no violation of the
B-CHSH inequality for standard (non-sequential) experiments.

\section{Local realistic description of sequences of measurements}

The  local hidden variable model for joint probabilities for
obtaining the results $a$ and $b$ upon the  performance of the
local measurements $A$, $B$, on an ensemble of pairs of particles in a
certain quantum-mechanical state,
must have the following structure
\begin{equation}
P_{A,B}(a,b)
=\int P_A(a;\lambda) \tilde P_B(b;\lambda)\varrho(\lambda)\text d \lambda,
\label{stand}
\end{equation}
where  $\lambda$ is the hidden
variable, $\varrho$ is its probability
distribution (and is independent of the choice of
$A$ and $B$), and
$P_A(a;\lambda)$ and $\tilde P_B(b;\lambda)$ are the probabilities
of
obtaining specific results, provided we measure the specified
observable ($A$ or $B$), and the element of the ensemble is in the hidden
variable state $\lambda$ (for further details, please consult \cite{CH}).
Usually, to show that statistics produced by quantum mechanical states cannot be
simulated by the above formula, $A$ and $B$ have been  treated as single
von Neumann measurements (represented by Hermitian operators).
%The set of
%results of such measurement is thus given by spectrum of the
%appropriate operator.
However quantum mechanics allows us to predict statistics of results
of much more
complicated  experiments. Consequently one should put
$A=\{A_1, \ldots A_{n_A}\}$ and $B=\{B_1, \ldots B_{n_B}\}$ where $A_i$, $B_i$
denote single generalized measurements (not only von Neumann ones),
with sets of results $I_{A_i}$ and $I_{B_i}$ respectively.
Then the formula (\ref{stand}) can be rewritten as
\begin{eqnarray}
&&P_{A_1,\ldots A_{i_A},B_1,\ldots B_{i_B}}(a_1,\ldots a_{i_A},
b_1,\ldots,b_{i_B})\nonumber\\
&&=\int P_{A_1,\ldots A_{i_A}}(a_1,\ldots a_{i_A};\lambda)
\tilde P_{B_1,\ldots B_{i_B}}(b_1,\ldots,b_{i_B};\lambda)
\varrho(\lambda)\text d \lambda,
\label{model}
\end{eqnarray}
where $P_{A_1,\ldots A_{i_A},B_1,\ldots B_{i_B}}(a_1,\ldots a_{i_A},
b_1,\ldots,b_{i_B})$ stands for the joint probability of obtaining outcomes
$\{a_1,\ldots a_{i_A}, b_1,\ldots,b_{i_B}\}$ in the measurements
$\{A_1,\ldots A_{i_A},B_1,\ldots B_{i_B}\}$. It should be emphasised here
that the measurements are performed on a {\it single} (but compound)
system which is a member of the ensemble. Now,
we can ask whether the statistics  predicted by quantum mechanics
can be reproduced by the above formula, i.e. whether the statistics
can be ascribed to the subsystems in a local-realistic way.
Below we will derive constraints implied by the existence of a the local
hidden variable model of the form (\ref{model}). These constraints can be
treated as the ones of Bell, albeit
generalized to the case of sequences of measurements.

For simplicity we will take into account an experiment consisting of two
consecutive measurements on each subsystem. The LHV model must
then have the form of
\begin{equation}
P_{A_1,A_2,B_1,B_2}(a_1,a_2,b_1,b_2)=
\int P_{A_1,A_2}(a_1,a_2;\lambda)\tilde P_{B_1,B_2}(b_1,b_2;\lambda)\varrho
(\lambda)
\text{d}\lambda.
\label{joint}
\end{equation}
Consider the
conditional probability
\begin{equation}
P_{A_1,A_2,B_1,B_2}(a_2,b_2|a_1,b_1)=
{P_{A_1,A_2,B_1,B_2}(a_1,a_2,b_1,b_2)\over
\sum_{a'_2\in I_{A_2}\atop b_2'\in I_{B_2}}
P_{A_1,A_2,B_1,B_2}(a_1,a'_2,b_1,b'_2)},
\label{cond}
\end{equation}
which is the
probability of obtaining outcomes $a_2$ and $b_2$ in the measurement
$A_2$, $B_2$ given that the measurement $A_1$, $B_1$ produced outcomes $a_1$,
$b_1$, respectively. Now we ask whether the form of the
formula (\ref{joint})
implies a similar one for the conditional probabilities (\ref{cond}).

Let
us introduce the following shorthand notation
\begin{equation}
P_{A_1,A_2,B_1,B_2}(a_1,b_1)=  \sum_{a'_2\in I_{A_2}\atop b_2'\in I_{B_2}}
P_{A_1,A_2,B_1,B_2}(a_1,a'_2,b_1,b'_2),
\end{equation}
and by	$P_{A_1,A_2}(a_1;\lambda)$ and
$\tilde P_{B_1,B_2}(b_1;\lambda)$ let us denote the marginals of
$P_{A_1,A_2}(a_1,a_2;\lambda)$ and  $\tilde P_{B_1,B_2}(b_1,b_2;\lambda)$,
respectively. Finally,	$P_{A_1,A_2}(a_1|a_2;\lambda)$,
$P_{B_1,B_2}(b_1|b_2;\lambda)$ are the appropriate conditional
probabilities. Therefore, one can write
\begin{equation}
P_{A_1,A_2,B_1,B_2}(a_2,b_2|a_1,b_1)=
\int
P_{A_1,A_2}(a_1|a_2;\lambda)
\tilde P_{B_1,B_2}(b_1|b_2;\lambda)
{P_{A_1,A_2}(a_1;\lambda) \tilde P_{B_1,B_2}(b_1;\lambda)\over
P_{A_1,A_2,B_1,B_2}(a_1,b_1)}\varrho(\lambda)\text{d}\lambda.
\end{equation}

Now, we observe that the conditional probability is given by the average
of the product $P_{A_1,A_2}(a_1|a_2;\lambda) P_{B_1,B_2}(b_1|b_2;\lambda)$
over the new probability distribution $\varrho_{A_1,A_2,B_1,B_2}(a_1,b_1;\lambda)$
defined as
\begin{equation}
\varrho_{A_1,A_2,B_1,B_2}(a_1,b_1;\lambda)=
{P_{A_1,A_2}(a_1;\lambda) \tilde P_{B_1,B_2}(b_1;\lambda)\over
P_{A_1,A_2,B_1,B_2}(a_1,b_1)}\varrho(\lambda).
\label{distr}
\end{equation}

Note, that what we have done so far is based only upon one
assumption of a
physical nature
contained in (\ref{joint}), while the rest consist of  merely mathematical
manipulations.
Now, we will use an argumentation
based upon the principles of local-realism \cite{pop2}: since the measurement
$A_2$, $(B_2)$ is performed after $A_1$, $(B_1)$, therefore the probabilities
$P_{A_1,A_2}(a_1;\lambda)$ and $P_{B_1,B_2}(b_1;\lambda)$ cannot depend on
$A_2$ and $B_2$, respectively. Further, one can put
$P_{A_1,A_2,B_1,B_2}(a_1,b_1)= P_{A_1,B_1}(a_1,b_1)$.
Otherwise, we would obtain violation of
causality. Thus one can drop the indices $A_2$ and $B_2$ in the distribution
(\ref{distr}). Now given  arbitrarily chosen measurements $A_1$, $B_1$ and
certain outcomes $a_1$, $b_1$, one can	denote by $X$
the full set of these conditions (i.e.,
$X\equiv\{A_1,B_1,a_1,b_1\}$).
Thus,  the conditional probabilities $P^X_{A_2,B_2}(a_2,b_2)\equiv
P_{A_1,A_2,B_1,B_2}(a_2,b_2|a_1,b_1)$ acquire the following
form
\begin{equation}
P^X_{A_2,B_2}(a_2,b_2)=\int
P^X_{A_2}(a_2;\lambda)P^X_{B_2}(b_2;\lambda)\varrho^X(\lambda)\text{d}\lambda,
\label{const}
\end{equation}
where $\varrho^X $ is a probability distribution, which is
{\it independent} of the
particular choice
of the measurements
$A_2$, $B_2$. As $X$ is independent of $A_2$, $B_2$,
and therefore also of $a_2$ and $b_2$,
 the conditional probabilities acquire the typical form
for the standard
local hidden variable models (\ref{stand}).
Obviously, they satisfy Bell's
inequalities.
This means that local realism
implies that any subensemble selected by local
measurements is describable by a LHV model.
Thus, if one can show
that according to quantum mechanics
such a sub-ensemble violates certain Bell's inequalities,
this implies that
the original state (for the whole ensemble)
does not allow any local-realistic
description of sequential measurements.
I.e., one can reveal the  impossibility of constructing the LHV model for joint
probabilities of sequences measurements indirectly, by checking that some
conditional probabilities do not admit the model. This task is much
easier,
as we can use standard Bell's inequalities, as proposed by Popescu \cite{pop2},
while in general we do not have
their counterpart for distributions of rank larger than
two.
%The using of standard Bell's inequalities in second measurement was
%first proposed by Popescu \cite{pop2}.
What is very important,
the subensemble is selected before the second measurement (of a Bell
observable). However, results of the second measurement
cannot be described in a local realistic way. Thus, as a subensemble
does not admit a LHV description, the full ensemble does not either.
{\it In simple words, the existence of a model given by (\ref{const})
is a necessary condition for the existence of model (\ref{joint}).}

The temporal sequence: first the pre-selection, later the actual
measurement of the Bell observable, is essential here.
Otherwise we would have
dealt with a problem equivalent to the one
associated with inefficient detection (the so-called {\it detection
loophole}).
Suppose for a while that the measurement $A_1$, $B_1$ is performed after
$A_2$, $B_2$ and the outcomes are simply detection or nondetection of the
particle. Then the probability (\ref{cond}) is the {\it aposteriori}
conditional
probability under the condition that both of the particles of the pair are
detected. Of course, now one cannot drop the indices $A_1$, $B_1$ in the
distribution (\ref{distr}). Hence, the condition (\ref{joint}) does not imply
similar constraints for  the aposteriori probabilities. In other words, if we
deal
with postselection, the hidden variable can contain the information how to
modify the probability of the outcome of the
second measurement according to what
observable was measured in the first one. In contrast, if the first
measurement is used for a pre-selection, its result is (in
accordance with local-realism) independent of
what is to be measured in future.
Otherwise, causality (locality) does not apply anymore.

With the above reasoning leading to (\ref{const}), we are now able to broaden the class of
known quantum states which have statistical properties which
violate  the assumptions of local
realism. In the
quantum case, the general measurement is described by a partition of unity
$\{V_i\}_{i=1}^n$ where $\sum V_i V_i^\dagger=I$, and
each $V_i$ corresponds
to a particular outcome, the probability of which, if the
system is in the state $\rho$, is
$p_i=\text{Tr}(V_i\rho V_i^\dagger)$. If the measurement produces
the outcome $V_i$ the system ends up in the state given by
\begin{equation}
\tilde\rho={V_i\rho V_i^\dagger\over \text{Tr}V_i\rho V_i^\dagger}.
\end{equation}
Thus for any state $\rho$ acting on Hilbert space ${{\cal H}_1}\otimes
{\cal H}_2$, if the state $\tilde\rho$ given by
\begin{equation}
\tilde\rho={V\otimes W\rho V^\dagger\otimes W^\dagger \over
\text{Tr}V\otimes W\rho V^\dagger\otimes W^\dagger},
\end{equation}
with  arbitrary  (bounded) operators $V$, $W$, does not admit the LHV model
for single von
Neumann measurements, then the state $\rho$ violates local
realism  in sequential measurements.
Indeed, one can take as $A_1$ and $B_1$ the partitions of unity given by
$\{\tilde V,\sqrt{I-\tilde V\tilde V^\dagger}\}$  and
$\{\tilde W,\sqrt{I-\tilde W\tilde W^\dagger}\}$ (where $\tilde V={V\over||V||}$
with $||V||$ being operator norm of $V$)
respectively. Here $\tilde V$ and $\tilde W$ should be associated
with the outcomes $a_1$ and
$b_1$.

Below we will study an experimental proposal aimed at revealing
empirically the non-existence
of  a local realistic model for a two photon state
which does not violate the usual B-CHSH
inequality. The scheme is based on the filtering method proposed by Gisin
\cite{Gis3}.

\section{Operational example}

In this section we shall present an experimental setup which can be used to
demonstrate violation of local realism in the case of sequences of
measurements. In other words, we aim at presenting an
operational discussion of the thesis presented earlier.

Throughout this section we shall employ solely standard techniques
of experimental quantum optics. As the primary sources we shall use
laser pumped non-linear crystals in which the phenomenon
of parametric down conversion leads to production of pairs of
entangled photons.

The exemplary initial state of the two-photon system has been suggested
in \cite{fazy} and is given by
\begin{equation}
\rho=\sum^2_{i=1}p_i|\psi_i\rangle\langle\psi_i|,
\end{equation}
with $p_1\neq p_2$,
and
\begin{eqnarray}
&|\psi_1\rangle=\alpha|2\rangle |2'\rangle +\beta|1\rangle |1'\rangle,&\\
&|\psi_2\rangle=\alpha|2\rangle |1'\rangle +\beta|1\rangle |2'\rangle,&
\end{eqnarray}
where,
we have used the convention that the first ket
describes the first subsystem, etc., and we have
$\langle1'|2'\rangle=\langle1|2\rangle=0$. The
coefficients $\alpha$ and $\beta$ are real and satisfy
\begin{equation}
(p_1-p_2)^2\leq(\alpha^2-\beta^2)^2.
\end{equation}
Such states do not violate the CHSH-Bell inequality.
Below,	it will be shown, in an operational
manner, that $\rho$ leads to
statistical correlations for sequential local experiments
that cannot be reproduced
by any local hidden variable theory.

First, let us concentrate on the question of actually
producing such states. To this end we propose to use two non-linear
crystals, as shown in fig.1. Both are pumped by a single laser. Its coherent
beam is split by a beamsplitter of reflectivity $|\alpha|^2$ and
transmittivity $|\beta|^2$.
At each crystal the spontaneous down conversion process can happen
(for a detailed description of this phenomenon, see e.g. \cite{Mandel}).
The two photon radiation of the pair
of crystals can be described by
\begin{equation}
|\psi\rangle=\alpha|2\rangle|2"\rangle+\beta|1\rangle|1"\rangle
\end{equation}
(for details,
consult the figure 1). The stable phase relation between the two
components of
$|\psi\rangle$	can be
obtained provided the optical paths linking the crystals with the
beamsplitter differ by much less than the coherence length of the
laser radiation.

The radiation in the modes $2''$ and $1''$ enters a Mach-Zehnder
interferometer. If one assumes that the beamsplitters of this device
are symmetric 50-50 ones, then when the relative phase shift between the arms
is $0$ the Mach-Zehnder interferometer acts effectively as a mirror, whereas
when the phase shift is $\pi$ it behaves like a perfectly transparent object.
By this we mean that in the first case the state $|2''\rangle$
is transformed into $|1'\rangle$ and $|1"\rangle$ into $|2'\rangle$
(fig. 1). In the
``transparent" mode the state $|2''\rangle$
changes into $|2'\rangle$ and $|1"\rangle$ into $|1'\rangle$.
The phase shift in the interferometer should change between
the two values very rapidly and in a stochastic manner.
This can be achieved by various mechanical, acusto-optical, or other methods.
In this way the output of the full device of fig. 1 is
described by the density
matrix $\rho$.

The usual procedure is to perform a Bell type experiment on the initial
two particle system. If one aims at an experiment in which each
photon is effectively describable by a two dimensional Hilbert space,
one can achieve this by placing on the way of each of the photons a
Mach- Zehnder interferometer (fig.2). This device enables one to perform any $U(2)$
transformation \cite{RECK}. However, in our case we aim at pre-selecting
the ensemble of photon pairs. To this end we place
a beamsplitter in the path $2$. One of its outputs is fed to a local
Mach-Zehnder device, whereas the other one is directed towards
a detector $D$. The full setup is presented in fig.3. To make the measurements
sequential in time the length of the optical path, $l$, between
the beamsplitter and the Mach-Zehnder interferometer
should satisfy $l\gg\Delta Tc$, where $\Delta T$ is the resolution
time of the detectors employed.

The beamsplitter $BS$ of fig.3 has a suitably chosen transmittivity
$|\beta/\alpha|^2$.
Also we assume it to be a symmetric device, which upon reflection
adds a phase shift of $\pi/2$. Thus the state $|2\rangle$ can be transformed
by BS into
\begin{equation}
(\beta/\alpha)|2\rangle + i\sqrt{1-(\beta/\alpha)^2}|D\rangle,
\end{equation}
where $|D\rangle$ denotes the state of a photon on its way to
the detector $D$.

The sub-ensemble of coincident counts behind {\it both} Mach-Zehnder
interferometers of fig.3 is effectively described by a new density matrix
which  reads
\begin{equation}
\rho' = \sum^2_{i=1}p_i|\psi'_i\rangle\langle\psi'_i|,
\end{equation}
where
\begin{eqnarray}
&|\psi'_1\rangle=\frac{1}{\sqrt{2}}(|2\rangle |2'\rangle +|1\rangle |1'\rangle)&\\
&|\psi'_2\rangle=\frac{1}{\sqrt{2}}(|2\rangle |1'\rangle +|1\rangle |2'\rangle).&
\end{eqnarray}
The mixed state described by $\rho$ violates the CHSH inequalities
(as shown by \cite{fazy}). Thus, via a local selection process
we get a subensemble of results which cannot be described
by any local hidden variable theory.
Therefore,  one can infer from the discussion presented in Section 2 that the initial
state $\rho$ (of the full ensemble) gives predictions for sequential
measurements which cannot have a local realistic interpretation.

Note added:
{\it After this paper was completed, we got the manuscripts of Popescu an Mermin
(Conference in Haifa, 1995). The authors present the results contained in the
second section (``Local realistic description of sequences of
measurements'') of our paper.}

M.\.Zukowski acknowledges partial support of the University of
Gda\'nsk research grant no. BW-5400-5-0087-6.

\begin{figure}
\caption{Generation of the required initial mixed state. Two
parametric down converters (PDC) are pumped by a single cw laser. The
beamsplitter BS has reflecitvity $|\alpha|^2$ and transmittivity $|\beta|^2$.
Both optical paths linking this beamsplitter with the PDC's differ by much
less than the coherence length of the laser radiation.
A Mach-Zehnder interferometer is placed to the right with respect to
the primary PDC sources ($M$ stands for mirror, $\phi$ is
the phase shift, the beamsplitters BS are symmetric $50-50$ ones).
The internal phase $\phi$ stochastically jumps
between the two values: $0$,  with probability $p_1$, and $\pi$, with
probability $p_2=1-p_1$. The resulting density matrix $\rho$
of the two-photon radiation is given in
the main text.}
\label{1}
\end{figure}
\begin{figure}
\caption{Mach-Zehner interferometer. The external and internal phase
shifters enable one to obtain any $U(2)$ transformation (modulo
certain overall phase shifts). Compare the previous figure.}
\label{1}
\end{figure}
\begin{figure}
\caption{The overall experimental configuration. The two Mach-Zechner
interferometers (MZ) enable
one to measure any dichotomic observables \protect\cite{RECK} (compare
fig.1). We insert a beamsplitter of trasmittivity $|\beta/\alpha|^2$
into the path $2$. Only those photons which do not activate
the detector $D$ can produce coincidences behind the two spatially
separated Mach-Zehnder interferometers. The time required for the
light to travel from BS to the left MZ interferometer should be
longer than the resolution time of the detectors.   }
\label{1}
\end{figure}

\end{document}